\documentclass[conference]{IEEEtran}
\usepackage{tikz}
\usepackage{hyperref}
\hypersetup{
  colorlinks  = true,
  urlcolor    = blue,
  linkcolor   = blue,
  citecolor   = blue 
}
\usepackage{pgfplots}
\usepackage{subcaption}
\usepackage{algorithm}

\usepackage{algpseudocodex}
\pgfplotsset{compat=1.18}
\usepackage{syntax}
\usepackage[utf8]{inputenc}
\usepackage[normalem]{ulem} 
\usepackage[most]{tcolorbox}
\usepackage{xcolor}
\usetikzlibrary{calc}
\usepackage{xspace}
\newcommand{\LlamaSmall}{Llama-3.1-8B\xspace}
\newcommand{\Gemma}{gemma-2-9b-it\xspace}
\newcommand{\Mistral}{Mistral-Small-24B\xspace}
\newcommand{\QwenMid}{Qwen-2.5-32b\xspace}
\newcommand{\LlamaLarge}{Llama-3.3-70B\xspace}
\newcommand{\QwenLarge}{Qwen-2.5-72b\xspace}
\usepackage{cleveref}
\usepackage{booktabs}
\usepackage{verbatim}
\usetikzlibrary{shapes.geometric, positioning, arrows.meta}
\def\nonterm#1{{\color{nontermcolor}\textlangle\textnormal{\emph{#1}}\textrangle}}
\renewcommand{\ulitleft}{\normalfont\ttfamily}
\renewcommand{\litleft}{\bgroup\color{termcolor}`\ulitleft}
\renewcommand{\litright}{\ulitright'\egroup}

\renewcommand{\syntleft}{\bgroup\color{nontermcolor}$\langle$\normalfont\itshape}
\renewcommand{\syntright}{$\rangle$\egroup}

{%
\end{grammar}%
\end{small}%
}

\usepackage{colortbl}
\usepackage{graphicx}

\definecolor{light-gray}{gray}{0.85}

\definecolor{nontermcolor}{rgb}{0.4, 0.05, 0.0} 
\definecolor{absnontermcolor}{rgb}{0.4, 0.5, 0.0} 
\definecolor{evokcolor}{rgb}{0.8, 0.05, 0.1}    
\definecolor{incorrectcolor}{rgb}{0.5, 0.0, 0.13}
\definecolor{incompletecolor}{rgb}{0.0, 0.0, 0.55}
\definecolor{validcolor}{rgb}{0.33, 0.42, 0.18} 
\definecolor{retcolor}{rgb}{0.65, 0.16, 0.16}
\definecolor{ipatterncolor}{rgb}{0.0, 0.5, 0.4} 
\definecolor{termcolor}{rgb}{0.0, 0.05, 0.4}    
\definecolor{regexcolor}{rgb}{0.1, 0.3, 0.1}    

\def\|#1|{\textit{#1}} 
\def\<#1>{\texttt{#1}} 
\newcommand{\ddmin}{\textit{ddmin}\xspace} 

\tcbset{
    promptbox/.style={
        sidebyside,
        sidebyside align=center,
        lefthand width=2.5cm, 
        colback=white,
        colframe=black,
        sharp corners,
        boxrule=1pt,
        fonttitle=\bfseries
    },
    kept/.style={colframe=green!60!black, colback=green!5},
    removed/.style={colframe=red!60!black, colback=red!5}
}

\newcommand\mytitle{How Much Prompt Is Enough? A Blackbox Minimization of Few-Shots in LLMs}

\newcommand{\framework}{\textsc{DD-FSM}\xspace}

\newtcolorbox{result}{
  colback=gray!5, colframe=gray!50,
  boxrule=0.4pt, arc=3pt,
  left=6pt, right=6pt, top=4pt, bottom=4pt,
  center, width=0.95\linewidth,
  fontupper=\small\itshape
}

\begin{document}
\title{\mytitle}
\author{
\IEEEauthorblockN{Ali Alfageeh}
\IEEEauthorblockA{\small University of Houston}
\and
\IEEEauthorblockN{Rahul Gopinath}
\IEEEauthorblockA{\small The University of Sydney}
\and
\IEEEauthorblockN{Amin Alipour}
\IEEEauthorblockA{\small University of Houston}
}

\IEEEpeerreviewmaketitle

\maketitle

\begin{abstract}
Prompts are the primary mechanism for directing the behavior of large language models (LLMs).
Yet the internal structure and causal hierarchy of prompts remain poorly understood:
which parts are causally necessary and which are redundant is an open question.
This opacity can have severe consequences.
Subtle prompt variations can silently shift model outputs in critical software systems,
and engineers lack techniques to reason about prompt reliability.

We present \framework, a blackbox prompt-minimization framework that
reduces few-shot prompts to their necessary minimal subset.
We use a case study to apply \framework to a few-shot learning system and demonstrate the insights that this framework can provide.
Our experiments show that few-shot exemplars can be reduced by a mean of 65.3\%~$\pm$~15.8\%
in character count while fully preserving propositional output fidelity.
The models preferentially retain
logical identifiers and constraint declarations while discarding natural
language prose and cross-prompt relational annotations.

Our analysis also shows that some models are universal encoders,
able to produce highly legible yet minimized prompts,
while others are universal decoders, able to interpret minimized prompts
from most other models.

By identifying which components are indispensable,
\framework provides a principled basis for prompt compression and structural analysis of few-shot exemplars.
\end{abstract}


\section{Introduction}
The integration of large language models (LLMs) into production software systems
has elevated the \emph{prompt} to a critical engineering artifact.
Two paradigms dominate inference-time control:
\emph{few-shot prompting}~\cite{brown2020},
in which a model is contextualized using example inputs and outputs,
and \emph{Chain-of-Thought (CoT) reasoning}~\cite{wei2022},
in which intermediate reasoning steps are explicitly demonstrated to elicit structured multi-step reasoning.
Industry analyses estimate that by 2025,
over 80\% of enterprises will incorporate prompt engineering into their workflows~\cite{gartner2024},
with few-shot strategies accounting for roughly 40\% of deployed techniques
and CoT variants comprising nearly a quarter of enterprise applications~\cite{sqmag2024}.
With over 67\% of organizations deploying LLM-powered products in production~\cite{indexdev2026},
prompts have become the cornerstone of software engineering.

Despite this scale, prompts are fragile.
LLMs are sensitive to superficial perturbations such as the ordering of examples~\cite{lu2022fantastically,zhao2021calibrate},
the phrasing of instructions,
and even the presence of seemingly irrelevant formatting tokens~\cite{min2022rethinking},
all of which can shift model outputs in ways invisible to developers.
Pearce et al.~\cite{pearce2022asleep} found that approximately 40\% of programs
generated by \mbox{GitHub} Copilot contained security vulnerabilities from the CWE Top 25,
illustrating that prompt-driven code generation can silently produce exploitable output.
Given the obscurity of the mapping from natural-language prompts to model behavior,
such failures can emerge at deployment without warning.

Current approaches to reliability assessment,
including static aggregate benchmarks such as MMLU and HumanEval
and qualitative human review,
treat the prompt–model interaction as a monolithic event.
As a result, they cannot answer a fundamental software engineering question:
\emph{which specific component of a prompt caused a given output?}
Recent work on automatic prompt optimization,
including DSPy~\cite{khattab2023dspy} and Automatic Prompt Engineering (APE)~\cite{zhou2022large},
addresses prompt \emph{construction} but not prompt \emph{explanation}:
these methods maximize task performance without revealing \emph{why} a prompt works
or \emph{which} of its components are causally necessary.
Without constraint-level observability,
engineering robust LLM applications remains a precarious trial-and-error process.
\begin{figure*}[tp]
\footnotesize
\tcbset{
    promptblock/.style={
        sidebyside,
        sidebyside align=top seam,
        lefthand width=2.6cm,
        colback=white,
        colframe=black,
        sharp corners,
        boxrule=0.8pt,
        fonttitle=\bfseries,
        left=4pt, right=4pt, top=3pt, bottom=3pt,
        before skip=2pt, after skip=2pt
    }
}

\newcommand{\removed}[1]{\textcolor{red!60!black}{\sout{#1}}}

\begin{tcolorbox}[promptblock, colback=red!5]
    \textbf{Role Assignment}
    \tcblower
    \removed{You are an assistant that analyzes Python prompts and generates propositional logic formalizations and relationships between prompts.}
\end{tcolorbox}

\begin{tcolorbox}[promptblock, colback=red!5]
    \textbf{Task Description}
    \tcblower
    \removed{Follow this example format strictly.}
\end{tcolorbox}

\begin{tcolorbox}[promptblock, colback=green!5]
    \textbf{Few-shot Examples}
    \tcblower
    \removed{\textbf{Prompts:}}

    \removed{\textbf{PromptID: P1}}\\
    \removed{Text:}
    \removed{Write a Python function named math(numbers\_input) that }
    takes a list of numbers as input and returns a new list consisting of the addition of each consecutive numbers in the input.\\
    C1: Python function\\
    C2: named math(numbers\_input)\\
    C3: takes a list of numbers as input\\
    C4: returns a new list\\
    C5: the addition\\
    C6: each consecutive numbers\\
    LogicalExpression: $P1 \rightarrow (C1 \land C2 \land C3 \land C4 \land C5 \land C6)$

    \vspace{2pt}
    \removed{\textbf{PromptID: P2}}\\
    \removed{Text:}
    \removed{Write a Python function named math(numbers\_input) that takes a list of numbers as input and returns a new array containing }
    of the product of each consecutive numbers in the input.\\
    C1: Python function\\
    C2: named math(numbers\_input)\\
    C3: takes a list of numbers as input\\
    C7: returns a new array\\
    C8: the \removed{product}\\
    \removed{C6: each consecutive numbers}\\
\removed{LogicalExpression: \textit{P2}  \textit{(C1  C2}} $\land C3 \land C7 \land C8 \land C6)$

    \vspace{2pt}
    \textbf{LogicalRelationshipWithPreviousPrompt:}

    PreviousPrompt: P1\\
    SemanticRefinement: C4 evolves into C7 from list to array, and C5 evolves into C8 from addition to product.\\
    NewAddition: \removed{Only refinements are introduced; no independent new constraints.\\
    CoreContinuation: $C1 \land C2 \land C3 \land C6$ remain unchanged.}
\end{tcolorbox}

\begin{tcolorbox}[promptblock, colback=green!5]
    \textbf{Constraints}
    \tcblower
    \removed{END OF EXAMPLE. Do not include the example in your response. Do} not generate any explanation more than the example structure. Only produce the formalization of the following two consecutive prompts. The logical expressions should be in C-form.
\end{tcolorbox}

\caption{Anatomy of a minimized prompt.
\colorbox{green!5}{\textcolor{green!60!black}{Green}} shading marks content retained after minimization;
\colorbox{red!5}{\textcolor{red!60!black}{red strikethrough}} marks removed spans.
The role assignment, task description, and post-example constraints are eliminated,
while the few-shot exemplar retains its logical skeleton of identifiers, constraints, and expressions.}
\label{fig:prompt-anatomy}
\end{figure*}

This paper proposes treating prompts as analyzable,
formally verifiable software artifacts.
We introduce \textbf{\framework} (\textbf{D}elta \textbf{D}ebugging for \textbf{F}ew-\textbf{S}hot \textbf{M}inimization),
a blackbox framework based on delta-debugging~\cite{zeller2002simplifying} that reduces prompts to their minimal necessary subset.

We use a few-shot system as a case study to demonstrate the analyses that this framework can produce to better understand and evaluate few-shot systems.
Input minimization requires a precise notion of behavioural preservation.
We capture this using \textbf{Prompt2Propositions (P2P)}~\cite{alfageeh2025from}, which maps prompts to propositional constraints
$P \equiv (C_1 \land \cdots \land C_k)$.
This allows direct equivalence checks between the baseline output and that of reduced prompts.
Minimization retains a reduction only if equivalence holds; otherwise, it is reverted. An example prompt and its minimization are given in \Cref{fig:prompt-anatomy}.



Experiments across six models and two temperature settings
show that few-shot exemplars can be reduced by a mean of 65.3\%~$\pm$~15.8\%
of characters while fully preserving propositional output fidelity.
The models preferentially retain
the logical skeleton of exemplars: prompt and constraint identifiers,
propositional declarations, and logical expressions, while discarding natural language prose
and cross-prompt relational annotations (\Cref{tab:ntStats}).
Regression analysis demonstrates that model scale is the dominant factor in achievable compression,
with each additional 10B parameters corresponding to 4.3 percentage points of additional removal.
Further, we find that some models 
serve as
\emph{universal encoders}, producing minimized prompts that retain
their fidelity when consumed by other models,
while others 
serve as \emph{universal decoders},
able to reconstruct the intended output from minimized prompts produced by
most other models, regardless of the degree of minimization.

\noindent\textbf{Contributions.}
\begin{enumerate}
\item \framework, a blackbox delta-debugging pipeline for few-shot prompt minimization leveraging an oracle based on propositional constraints.
\item An empirical characterization of which prompt components survive minimization, attributing retained and removed content to specific nonterminals of the P2P grammar.
\item A regression analysis that separates the effects of model scale, sampling
      temperature, and architecture family on minimization,
      showing that family-level variation dominates raw parameter count.
\item A cross-model transfer study that identifies
      \emph{universal encoders} (models whose minimized prompts travel well
      to other models) and \emph{universal decoders}
      (models that reliably consume minimized prompts produced by other models).
\end{enumerate}

The rest of the paper is organized as follows. \Cref{sec:motivation} provides the motivation and research questions. \Cref{sec:design} discusses the design and implementation. \Cref{sec:results} presents the results, followed by \Cref{sec:discussion}, which discusses the implications.
\Cref{sec:related} covers related work.
\Cref{sec:threats} discusses threats to validity, and \Cref{sec:conclusion} concludes.

\section{Motivation and Research Questions}\label{sec:motivation}
Existing work on few-shot prompt optimization and example compression often relies on heuristic pruning, embedding similarity, or end-to-end LLM evaluation~\cite{aftab2024tailored,shi2024prompt}. Although these approaches can shorten prompts while preserving task performance, they provide limited insight into what semantic content survives compression, why certain tokens remain indispensable, or how task intent persists even as symbolic structure collapses. This lack of interpretability limits critical understanding of in-context learning and weakens reproducibility across model runs.

Our \textbf{Delta Debugging for Few-Shot Minimization (DD-FSM)} framework tackles this by applying delta debugging directly to prompts to find the smallest subset that preserves the original output. This approach reveals what essential information survives as formatting and peripheral text are removed.
\framework gives a clear method for understanding how task intent survives during significant prompt reduction.

\subsection{Research Questions}
Prompt minimization is only useful if reduction is achievable without sacrificing output fidelity, and the reduction possible quantifies the redundancy present in typical prompts.
Hence we ask:\\
\noindent\textbf{RQ1.} To what extent can a prompt be minimized while fully preserving propositional output fidelity?

For insight into in-context learning, we need to know what is being preserved.
The types of content that survive minimization reveal which structural or semantic elements
LLMs treat as indispensable, and which are genuinely redundant.
Hence, we ask:\\
\noindent\textbf{RQ2.} Which parts of prompts survive minimization?

Preserving propositional output fidelity is a formal criterion,
but practitioners care about downstream task accuracy.
A minimized prompt that satisfies the oracle but degrades real-world performance
would have limited practical value. Hence, our question is:\\
\noindent\textbf{RQ3.} How does the task accuracy of the minimized prompt compare to that of the full prompt?

A minimized prompt derived from one model may have exploited model-specific
behaviors rather than identifying genuinely redundant content.
Testing transfer across models determines whether the minimal core
is universal or model-dependent. This is what we verify next.\\
\noindent\textbf{RQ4.} Does the minimized prompt preserve correctness across large language models?

\begin{figure*}[tbp]
\begin{grammar}
<PromptSeq> $\rightarrow$ <Text> `Prompts:' <Prompt>* 
`END OF EXAMPLE' <Text>

 
<Prompt> $\rightarrow$ `PromptID:' <PID> `Text:' <Text> <Constraint>*  `LogicalExpression:' <LExp> [<LRel>]


 
 
<LExp> $\rightarrow$ <PID> `->' `(' <Conjuncts> `)'

<LRel> $\rightarrow$ `LogicalRelationshipWithPreviousPrompt:' \\
       `PreviousPrompt:' <PID>  `Refinement:' <Text>  `Addition:' <Text>  `CoreContinuation:' <Conjuncts>
       
<Conjuncts> $\rightarrow$ <CID> $\wedge$ <Conjuncts> | <CID>
 
<Constraint> $\rightarrow$ <CID> `:' <Text>
 
<PID> $\rightarrow$ `P1' | `P2' | `P3' | \ldots

<CID> $\rightarrow$  `C1' | `C2' | `C3' | \ldots
\end{grammar}
\caption{Grammar for prompt minimization formalization (P2P)}
\label{fig:p2pGrammar}
\end{figure*}

\section{Design and Implementation}\label{sec:design}
We first provide an overview of the algorithm,
how we adapt it into the \framework pipeline,
and the models and dataset used in the evaluation.

\noindent\textbf{Delta Debugging.}
Zeller and Hildebrandt~\cite{zeller2002simplifying}
introduced the minimizing delta debugging algorithm \ddmin
(\Cref{alg:ddmin}).
It takes a predicate and a satisfying starting input
and iteratively simplifies the input
until it reaches the minimal input that still satisfies the predicate.
\begin{figure*}[tp]
    \centering
    \begin{tikzpicture}[
    >=Stealth,
    thick,
    rounded corners=4pt,
    rnd/.style={rectangle, rounded corners=5pt, draw,
                minimum width=3cm, minimum height=0.9cm,
                align=center, font=\small},
    stad/.style={rectangle, rounded corners=0.4cm, draw,
                minimum width=3cm, minimum height=0.8cm,
                align=center, font=\small},
    box/.style={rectangle, draw,
                minimum width=3cm, minimum height=0.9cm,
                align=center, font=\small},
    dec/.style={diamond, draw, aspect=1.8,
                align=center, font=\small, inner sep=2pt},
]
\node[dec]  (check)  at ( 0,    1.0) {Same\\Propositions?};
\node[box]  (reject) at ( 4.5,  2.0) {Reject Reduction\\(PASS)};
\node[box]  (accept) at ( 4.5,  0.0) {Accept Reduction\\(FAIL)};
\node[stad] (start)  at ( 4.5, -2.0) {Start: Full Prompt\\($C^{\text{full}}$)};
\node[rnd]  (dd)     at ( 9,    0)   {DD Module\\[3pt]$C' = C^{\text{full}} - i$};
\node[rnd]  (oracle) at (13,    2.0) {LLM Oracle\\Evaluate Candidate $C'$};
\node[stad] (stop)   at (13,    0)   {Stop: Minimal\\Correct Prompt};

\coordinate (bus) at (7.0, 0);

\draw[->, green!60!black]
    ([yshift=0.2cm]check.east) -- node[above, font=\small\itshape] {No} ++(0.5,0) |- (reject.west);
\draw[->, red!60!black]
    ([yshift=-0.2cm]check.east) -- node[below, font=\small\itshape] {Yes} ++(0.5,0) |- (accept.west);

\draw[->] (reject.east) -- (reject.east -| bus) -- ([yshift= 0.25cm]bus) -- ([yshift= 0.25cm]dd.west);
\draw[->] (accept.east) --                                                    (dd.west);
\draw[->] (start.east)  -- (start.east  -| bus) -- ([yshift=-0.25cm]bus) -- ([yshift=-0.25cm]dd.west);

\draw[->] (dd.north) |-  (oracle.west);
\draw[->] (dd.east)  --  (stop.west);

\draw[->] (oracle.north) -- ++(0,0.6) -| (check.north);

\end{tikzpicture}%
    \caption{\framework workflow for prompt minimization and evaluation.}
    \label{fig:workflow}
\end{figure*}
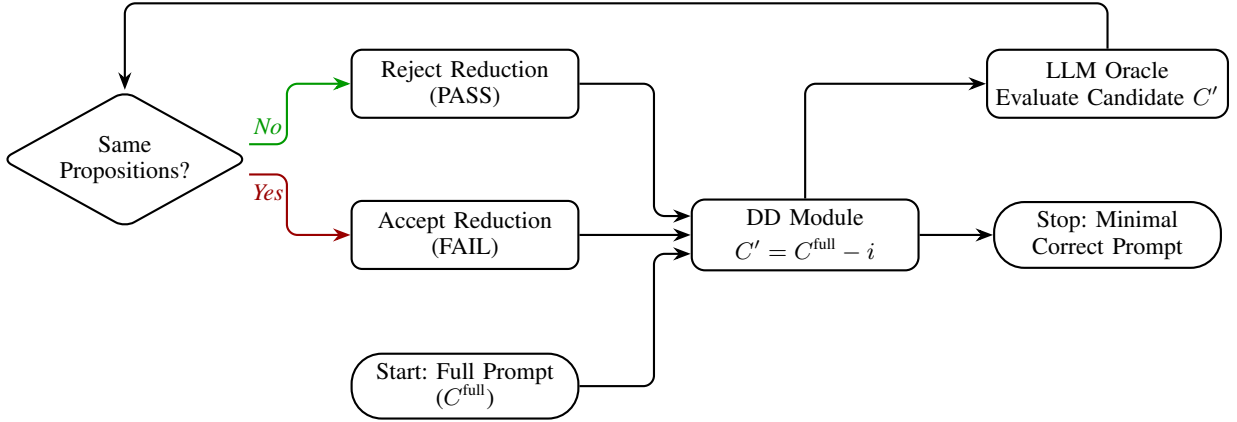
\begin{algorithm}
\caption{Delta Debugging (ddmin)}
\label{alg:ddmin}
\begin{algorithmic}
\Require $C$ such that $\Call{Test}{C} = \text{FAIL}$
\Ensure $C'$ is 1-minimal failing subset
\State $n \gets 2$ \Comment{Initial granularity}
\While{$|C| \ge 2$}
    \State Partition $C$ into $n$ segments $\Delta_1, \Delta_2, \dots, \Delta_n$
    \State $some\_reduction \gets \text{false}$
   \ForAll{$\Delta_i$}
        \If{$test(\Delta_i) = \text{FAIL}$}
            \State $C \gets \Delta_i$
            \State $n \gets 2$
            \State $some\_reduction \gets \text{true}$
            \State \textbf{break}
        \EndIf
    \EndFor
    \If{not $some\_reduction$}
        \ForAll{$\Delta_i$}
            \If{$\Call{Test}{C \setminus \Delta_i} = \text{FAIL}$}
                \State $C \gets C \setminus \Delta_i$
                \State $n \gets \max(n - 1, 2)$
                \State $some\_reduction \gets \text{true}$
                \State \textbf{break}
            \EndIf
        \EndFor
    \EndIf

    \If{not $some\_reduction$}
        \If{$n < |C|$}
            \State $n \gets \min(|C|, 2n)$
        \Else
            \State \textbf{break}\Comment{Result is 1-minimal}
        \EndIf
    \EndIf
\EndWhile
\Return $C$
\end{algorithmic}
\end{algorithm}

To identify which parts of the few-shot exemplar guide the model to reproduce
the same propositional output as the manually authored baseline,
we integrate \ddmin into \framework for few-shot exemplar reduction,
as described below.

\noindent\textbf{Workflow.} \Cref{fig:workflow} shows a high-level overview of the \framework workflow.

\noindent\textbf{Framework.} The framework targets the setting studied by Alfageeh et al.~\cite{alfageeh2025from}, who examined the interaction between students and LLMs on programming tasks using logical representations. The logical representations were generated by the LLM using few-shot examples. They found that 98\% of the sample correctly mapped the student's prompts to the logical representations. We therefore seek to understand how the model generated these logical representations using the few-shot examples provided to the LLM.

\begin{enumerate}
\item \textbf{Ground Truth.} In this setup, we manually coded two inquiries, each inquiry having two prompts, into logical expressions. These logical expressions serve as an initial check for the model when it uses the full few-shot examples to generate the baselines for each inquiry.
\item \textbf{Delta Debugging.} In this step, \ddmin begins minimizing the full few-shot example. The minimized few-shot example is sent to the model, followed by the two inquiries. The minimized few-shot is categorized into one of the outcome classes discussed in \Cref{subsec:dd-verdicts}.
\item \textbf{Verification.} Lastly, we verify the minimal example returned by \ddmin for further analysis.
\end{enumerate}

\subsection{Evaluated Models}
To broaden the generalizability of our findings, we selected six large language models spanning four model families and parameter counts from 8B to 72B. We evaluate each model at two temperature settings (\(T=0.0\) and \(T=0.5\)). The models are listed in \Cref{tab:modelsused}.
\begin{table}[ht]
\centering
\caption{Large Language Models Used in Evaluation.}
\label{tab:modelsused}
\renewcommand{\arraystretch}{1.2}
\begin{tabular}{|l|l|c|}
\hline
\textbf{Model Name} & \textbf{Family} & \textbf{Parameters} \\
\hline\hline
\LlamaSmall{}      & Llama  & 8B \\
\Gemma{}     & Gemma  & 9B \\
\Mistral{} & Mistral & 24B \\
\QwenMid{}      & Qwen   & 32B \\
\LlamaLarge{}     & Llama  & 70B \\
\QwenLarge{}      & Qwen   & 72B \\
\hline
\end{tabular}
\end{table}

\subsection{Analysis Instrumentation}\label{subsec:instrumentation}

\framework records the result of each minimization run as a JSON parse tree
in which every internal node is labeled with its P2P grammar nonterminal
(see \Cref{fig:p2pGrammar})
and every leaf is either a retained terminal string
or a deletion marker. 
After minimization is complete,
a post-hoc analysis script traverses this tree recursively:
for each nonterminal $N$ it accumulates
(i) the total character count
of all terminals in the subtree rooted at $N$,
and (ii) the count of characters carried by deletion markers in that subtree.
The ratio of deleted to total characters gives the removal rate for $N$.
Aggregating these rates across all 12 model/temperature runs
yields the mean and standard deviation reported in \Cref{tab:ntStats}.

Because the P2P grammar is hierarchical,
a character deleted inside a \nonterm{Constraint} subtree
is also counted under every ancestor---\nonterm{ConstraintList},
\nonterm{Prompt}, and \nonterm{PromptSeq}.
The per-nonterminal rates therefore reflect the aggregate behavior
of all content governed by that symbol, not just its immediate children.

\subsection{Verdict Classification in Delta Debugging}
\label{subsec:dd-verdicts}

The framework classifies each minimization step into one of three verdicts.
These verdicts reflect how the Delta Debugging (DD) algorithm interprets the behavior of a minimized few-shot exemplar relative to predefined baselines.
Note that in Zeller's original formulation, the verdict names \texttt{FAIL}/\texttt{PASS} carry their conventional software-testing meaning (a \texttt{FAIL} means the bug is still triggered).
To avoid confusion in the prompt-minimization context, we adopt the aliases
\texttt{PRESERVED}, \texttt{LOST}, and \texttt{UNRESOLVED} throughout this paper.

\subsubsection{Preserved State: Target Behavior Retained}

A \texttt{PRESERVED} outcome (corresponding to \texttt{FAIL} in classical DD) indicates that the reduced input still \emph{preserves} the target behavior.

\begin{enumerate}
    \item \textbf{PRESERVED (doubly verified match)}:  
    The reduced prompt successfully generates logical expressions that match the baselines for both Inquiry~1 and Inquiry~2. Additionally, the result is fully deterministic: when re-evaluated during Verification, the same matching outputs are reproduced.
    
    \item \textbf{DD Interpretation}:  
    This outcome is treated as \texttt{PRESERVED}, meaning the target property is retained. The algorithm continues minimizing the input.
\end{enumerate}

\subsubsection{Lost States: Behavior No Longer Preserved}

A \texttt{LOST} outcome (corresponding to \texttt{PASS} in classical DD) indicates that the minimized prompt no longer preserves the target behavior. These cases typically arise when the prompt becomes too degraded for reliable parsing.

\begin{enumerate}
    \item \textbf{LOST (\textless 2 expressions Inq1)}:  
    For the first run, the model fails to produce exactly two structured expressions.

    \item \textbf{LOST (\textless 2 expressions Inq2)}:  
    Inquiry~1 succeeds, but Inquiry~2 fails to produce two valid expressions.

    \item \textbf{DD Interpretation}:  
    Both cases are treated as \texttt{LOST}, indicating the absence of the target property. The configuration is discarded.
\end{enumerate}

\subsubsection{Ambiguous States: Divergent or Non-Deterministic Behavior}

An \texttt{UNRESOLVED} outcome indicates that the model produces outputs, but they either deviate from the baseline or exhibit instability.

\noindent\textbf{Divergent Outputs.}
\begin{enumerate}
    \item \textbf{LOST (diff Inq1)}:  
    Inquiry~1 produces valid expressions that do not match the baseline.

    \item \textbf{LOST (diff Inq2)}:  
    Inquiry~1 matches the baseline, but Inquiry~2 produces non-matching expressions.

    \item \textbf{DD Interpretation}:  
    Although labeled as \texttt{LOST} in logs, these cases are internally treated as \texttt{UNRESOLVED}.
\end{enumerate}

\noindent\textbf{Flaky Behavior}
If a configuration initially matches both baselines but fails to do so consistently during verification, it is classified as flaky:

\begin{enumerate}
    \item \textbf{UNRESOLVED -- Flaky Inq1 (\textless 2 expressions Inq1 Verif)}:  
    Re-evaluation of Inquiry~1 fails to extract valid expressions.

    \item \textbf{UNRESOLVED -- Flaky Inq1 (diff Inq1 Verif)}:  
    Re-evaluation of Inquiry~1 produces different expressions.

    \item \textbf{UNRESOLVED -- Flaky Inq2 (\textless 2 expressions Inq2 Verif)}:  
    Re-evaluation of Inquiry~2 fails to extract valid expressions.

    \item \textbf{UNRESOLVED -- Flaky Inq2 (diff Inq2 Verif)}:  
    Re-evaluation of Inquiry~2 produces different expressions.

    \item \textbf{DD Interpretation}:  
    All such cases are treated as \texttt{UNRESOLVED}, indicating instability or lack of reproducibility.
\end{enumerate}

\section{Results}\label{sec:results}
\begin{table}[tbp]
\centering
\caption{Removal statistics by model and sampling temperature.}
\label{tab:modelremoved}
\renewcommand{\arraystretch}{1.2}
\begin{tabular}{|l|c|r|r|r|}
\hline
\textbf{Model} & \textbf{T} & \textbf{Total} & \textbf{Removed} & \textbf{Removed (\%)} \\
\hline\hline
\Gemma{}     & 0.0 & 1749 & 1110 & 63.5 \\
\Gemma{}     & 0.5 & 1749 & 1273 & 72.8 \\
\hline
\LlamaSmall{}      & 0.0 & 1749 & 844  & 48.3 \\
\LlamaSmall{}      & 0.5 & 1749 & 964  & 55.1 \\
\hline
\Mistral{} & 0.0 & 1749 & 672  & 38.4 \\
\Mistral{} & 0.5 & 1749 & 869  & 49.7 \\
\hline
\QwenMid{}      & 0.0 & 1749 & 1231 & 70.4 \\
\QwenMid{}      & 0.5 & 1749 & 975  & 55.7 \\
\hline
\LlamaLarge{}     & 0.0 & 1749 & 1526 & 87.3 \\
\LlamaLarge{}     & 0.5 & 1749 & 1474 & 84.3 \\
\hline
\QwenLarge{}      & 0.0 & 1749 & 1396 & 79.8 \\
\QwenLarge{}      & 0.5 & 1749 & 1372 & 78.4 \\
\hline
\hline
\textbf{Average}  & ---   & 1749 & \textbf{1142} & \textbf{65.3} \\
\textbf{St.\ Dev.} & ---  & ---   & \textbf{265}  & \textbf{15.8} \\
\hline
\end{tabular}
\end{table}

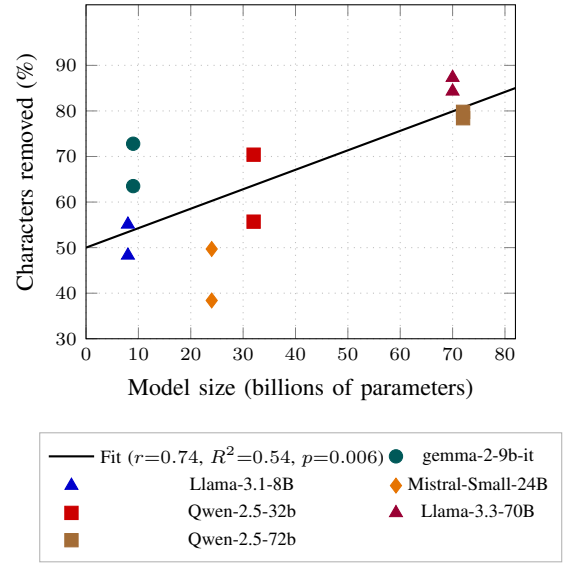
\begin{figure}[tp]
  \centering
  \begin{tikzpicture}
\begin{axis}[
    width=0.82\columnwidth,
    height=6cm,
    enlarge y limits={abs=0.5cm, upper},
    xlabel={Model size (billions of parameters)},
    ylabel={Characters removed (\%)},
    xmin=0, xmax=82,
    ymin=30, ymax=95,
    xtick={0,10,20,30,40,50,60,70,80},
    ytick={30,40,50,60,70,80,90},
    grid=major,
    grid style={dotted, gray!50},
    legend style={at={(0.5,-0.28)}, anchor=north, font=\scriptsize,
                  draw=gray!60, fill=white, inner sep=3pt,
                  legend columns=2},
    tick label style={font=\scriptsize},
    label style={font=\small},
]

\addplot[thick, black, domain=0:82, samples=2] {0.4271*x + 50.00};
\addlegendentry{Fit ($r{=}0.74$, $R^2{=}0.54$, $p{=}0.006$)}

\addplot[only marks, mark=*, mark size=2.5pt, color=teal!70!black]
coordinates { (9, 63.5) (9, 72.8) };
\addlegendentry{\Gemma{}}

\addplot[only marks, mark=triangle*, mark size=2.8pt, color=blue!80!black]
coordinates { (8, 48.3) (8, 55.1) };
\addlegendentry{\LlamaSmall{}}

\addplot[only marks, mark=diamond*, mark size=2.8pt, color=orange!90!black]
coordinates { (24, 38.4) (24, 49.7) };
\addlegendentry{\Mistral{}}

\addplot[only marks, mark=square*, mark size=2.5pt, color=red!80!black]
coordinates { (32, 70.4) (32, 55.7) };
\addlegendentry{\QwenMid{}}

\addplot[only marks, mark=triangle*, mark size=2.8pt, color=purple!80!black]
coordinates { (70, 87.3) (70, 84.3) };
\addlegendentry{\LlamaLarge{}}

\addplot[only marks, mark=square*, mark size=2.5pt, color=brown!85!black]
coordinates { (72, 79.8) (72, 78.4) };
\addlegendentry{\QwenLarge{}}

\end{axis}
\end{tikzpicture}
  \caption{Scatter plot of achievable compression (\% characters removed)
           against model size (billions of parameters) for all 12 model/temperature runs.
           The line shows the ordinary least-squares fit
           ($y = 0.43x + 50.0$, $r = 0.74$, $R^2 = 0.54$, $p = 0.006$).
           Each model contributes two points (temperatures $T=0.0$ and $T=0.5$).}
  \label{fig:regressionPlot}
\end{figure}

\subsection{RQ1: Extent of Minimization}

\Cref{tab:modelremoved} shows the proportion of characters removed by each model
at each sampling temperature. 
Every model achieves substantial reduction while fully preserving propositional output fidelity,
confirming that typical few-shot prompts carry significant redundancy.
Across all 12 model/temperature combinations the mean removal is 65.3\%~$\pm$~15.8\%,
with individual runs ranging from 38.4\% (\Mistral{}, $T=0.0$)
to 87.3\% (\LlamaLarge{}, $T=0.0$).

Model parameter size seems to be the dominant factor in how much can be removed.
The two 70B-class models---\LlamaLarge{} and \QwenLarge{}---average 85.8\% and 79.1\% removal,
respectively,
whereas the 8B-class models achieve far less reduction:
\LlamaSmall{} averages 51.7\% and \Mistral{} averages 44.1\%. \QwenMid{} and \Gemma{} fall in between at 63.1\% and 68.2\%.
This gradient suggests that larger models maintain propositional fidelity
even when the scaffolding around the core examples is stripped away,
while smaller models rely more heavily on the examples and their surrounding structure.

Sampling temperature has a comparatively modest and inconsistent effect.
For \Mistral{}, raising \(T\) from 0.0 to 0.5 increases removal by 11.3 percentage points,
while for \QwenMid{} the same increase in temperature reduces removal by 14.7 points.
No systematic direction is apparent,
implying that the achievable compression is governed primarily by model capacity
rather than by stochasticity in decoding.

\begin{result}
\textbf{RQ1.}
\framework reduces prompts by an average of 65.3\% of characters
while fully preserving propositional output fidelity.
Reduction scales with model size:
70B-class models tolerate up to 87\% removal,
whereas 8B-class models plateau near 48--55\%.
Temperature has no consistent directional effect on compression.
\end{result}

\subsection{RQ2: What Prompt Content Survives?}

\begin{table}[tp]
  \centering
  \caption{Characters removed (\%) per nonterminal symbol}
\renewcommand{\arraystretch}{1.4} 
\begin{tabular}{|l|r|r|}
\hline
\textbf{Nonterminal} & \textbf{Mean (\%)} & \textbf{Std. Dev.} \\
\hline\hline
\texttt{\nonterm{PromptSeq}}      & 65.3 & 15.8 \\
\texttt{\nonterm{Prompt}}         & 60.7 & 19.5 \\
\texttt{\nonterm{Constraint}}     & 45.5 & 22.7 \\
\texttt{\nonterm{Conjuncts}}      & 61.1 & 27.7 \\
\texttt{\nonterm{LExp}}           & 50.6 & 22.9 \\
\texttt{\nonterm{LRel}}           & 72.6 & 26.3 \\
\texttt{\nonterm{PID}}            & 42.6 & 17.1 \\
\texttt{\nonterm{CID}}            & 53.6 & 16.7 \\
\hline
\end{tabular}
\label{tab:ntStats}
\end{table}

\Cref{tab:ntStats} breaks the removal rate down by nonterminal symbol in the P2P grammar.

\noindent\textbf{Relational annotations are the most removable content.}
Among the grammar symbols in \Cref{tab:ntStats},
\nonterm{LRel} (72.6\%~$\pm$~26.3\%) is the most heavily pruned.
\nonterm{LRel} encodes the logical relationship between consecutive prompts
({\small ``\texttt{LogicalRelationshipWithPreviousPrompt}''});
its high removal rate indicates that models do not rely on explicit cross-prompt relationship annotations
to reproduce correct propositional output.

\noindent\textbf{The logical skeleton is preserved.}
At the other end of the spectrum,
prompt and constraint identifiers survive far more often.
\nonterm{PID} (prompt identifiers such as P1, P2) is removed at only 42.6\%~$\pm$~17.1\%,
and \nonterm{CID} (constraint identifiers such as C1, C2) at 53.6\%~$\pm$~16.7\%.
\nonterm{Constraint} text (45.5\%~$\pm$~22.7\%) and \nonterm{ConstraintList} (45.9\%~$\pm$~24.7\%)
are also relatively preserved,
as is the \nonterm{LExp} logical expression structure (50.6\%~$\pm$~22.9\%).
Together, these nonterminals form the logical skeleton of the few-shot exemplar.


\begin{result}
\textbf{RQ2.}
Minimization preferentially eliminates natural language prose and cross-prompt relational annotations
while retaining the logical skeleton.
\end{result}

\subsection{RQ3: Task Accuracy of the Minimized Prompt}

To answer RQ3, we generated 10 independent responses
for each model using its minimized prompt
and compared the resulting constraint sets
against the baselines obtained from the full prompt.
The diagonals in \Cref{fig:crossModel} summarizes accuracy per model. 

Considering T=0.0, four models achieved 100\% accuracy across all 10 runs,
reproducing exactly the baseline constraints on both inquiries:
\Gemma{}, \QwenMid{}, \LlamaLarge{}, and \LlamaSmall{}.
The remaining two models exhibited distinct failure modes.
Considering T=0.5, only three models achieved 100\% accuracy: \Gemma{}, \Mistral{}, and \QwenLarge{}.

\noindent\textbf{\QwenLarge{}.}
For Inquiry~1 the model produced the correct number of constraints,
but the logical expression was rendered in free-form prose
rather than the required \(\rightarrow, \land\) notation;
we count such outputs as correct.
For Inquiry~2, under the full prompt the model produced
5 constraints for Prompt~1 and 6 for Prompt~2,
whereas under the minimized prompt it produced only 4 and 5, respectively.

\noindent\textbf{\Mistral{}.}
The model had a single failure mode:
for Inquiry~1 it generated a logical expression for Prompt~2 in only one of ten runs.
All other runs matched the baseline.

Overall,
four of six models reproduce the full-prompt behavior exactly
after minimization,
with output-format post-processing.

\begin{result}
\textbf{RQ3.}
Four of six models reproduce full-prompt task accuracy exactly
under their minimized prompts.
The remaining two show specific, interpretable failure modes such as inconsistent output formatting,
or omitted logical expressions
rather than a general collapse in accuracy.
\end{result}

\subsection{RQ4: Cross-Model Transfer of the Minimized Prompt}

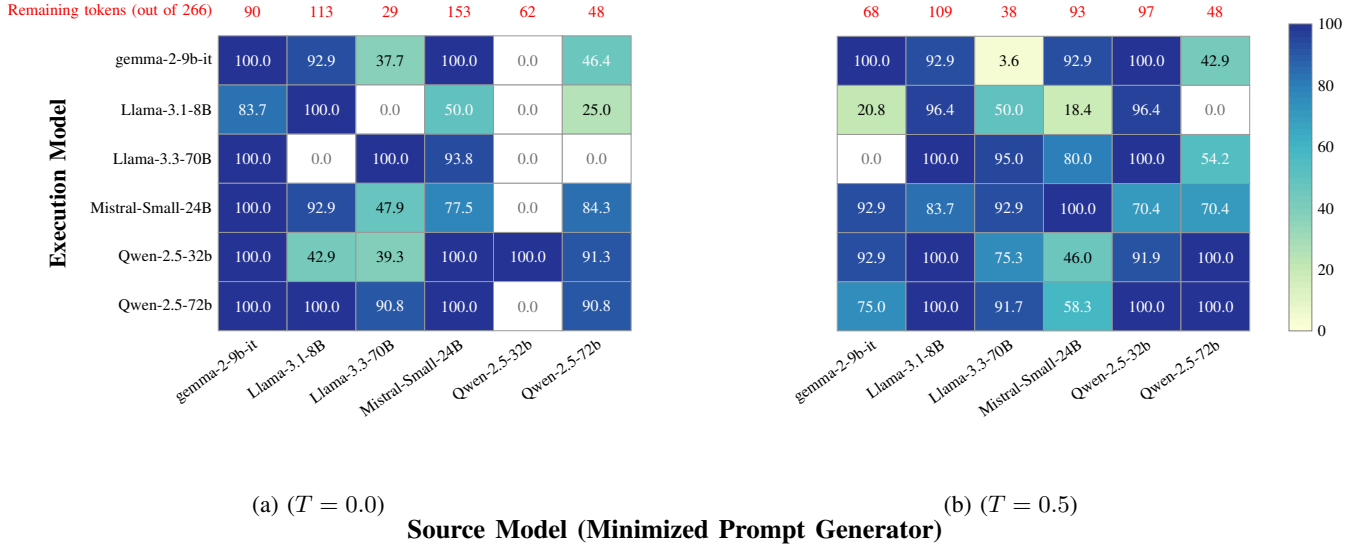
\begin{figure*}[tp]
    \centering
    \begin{subfigure}[t]{0.48\linewidth}
        \centering
        \resizebox{\linewidth}{!}{
            \newcommand{\unknown}{?.?}
\begin{tikzpicture}[
    every node/.style={font=\scriptsize},
    cell/.style={draw=black!40, line width=0.3pt, minimum width=1.05cm, minimum height=0.75cm, inner sep=0pt, outer sep=0pt, anchor=center},
  ]
  \definecolor{cmA}{RGB}{255,255,217}
  \definecolor{cmB}{RGB}{199,233,180}
  \definecolor{cmC}{RGB}{127,205,187}
  \definecolor{cmD}{RGB}{65,182,196}
  \definecolor{cmE}{RGB}{44,127,184}
  \definecolor{cmF}{RGB}{37,52,148}
  \node[anchor=center, inner sep=1pt] at (0.000,0.750) {\textcolor{red}{90}};
  \node[anchor=center, inner sep=1pt] at (1.050,0.750) {\textcolor{red}{113}};
  \node[anchor=center, inner sep=1pt] at (2.100,0.750) {\textcolor{red}{29}};
  \node[anchor=center, inner sep=1pt] at (3.150,0.750) {\textcolor{red}{153}};
  \node[anchor=center, inner sep=1pt] at (4.200,0.750) {\textcolor{red}{62}};
  \node[anchor=center, inner sep=1pt] at (5.250,0.750) {\textcolor{red}{48}};
  \node[anchor=east, inner sep=1pt] at (-0.575,0.750) {\textcolor{red}{Remaining tokens (out of 266)}};
  
  \node[cell, fill=cmE!0!cmF] at (0.000,0.000) {\textcolor{white}{100.0}};
  \node[cell, fill=cmE!35!cmF] at (1.050,0.000) {\textcolor{white}{92.9}};
  \node[cell, fill=cmB!12!cmC] at (2.100,0.000) {\textcolor{black}{37.7}};
  \node[cell, fill=cmE!0!cmF] at (3.150,0.000) {\textcolor{white}{100.0}};
  \node[cell, fill=white] at (4.200,0.000) {\textcolor{black!55}{0.0}};
  \node[cell, fill=cmC!68!cmD] at (5.250,0.000) {\textcolor{white}{46.4}};
  
  \node[cell, fill=cmE!82!cmF] at (0.000,-0.750) {\textcolor{white}{83.7}};
  \node[cell, fill=cmE!0!cmF] at (1.050,-0.750) {\textcolor{white}{100.0}};
  \node[cell, fill=white] at (2.100,-0.750) {\textcolor{black!55}{0.0}};
  \node[cell, fill=cmC!50!cmD] at (3.150,-0.750) {\textcolor{white}{50.0}};
  \node[cell, fill=white] at (4.200,-0.750) {\textcolor{black!55}{0.0}};
  \node[cell, fill=cmB!75!cmC] at (5.250,-0.750) {\textcolor{black}{25.0}};
  
  \node[cell, fill=cmE!0!cmF] at (0.000,-1.500) {\textcolor{white}{100.0}};
  \node[cell, fill=white] at (1.050,-1.500) {\textcolor{black!55}{0.0}};
  \node[cell, fill=cmE!0!cmF] at (2.100,-1.500) {\textcolor{white}{100.0}};
  \node[cell, fill=cmE!31!cmF] at (3.150,-1.500) {\textcolor{white}{93.8}};
  \node[cell, fill=white] at (4.200,-1.500) {\textcolor{black!55}{0.0}};
  \node[cell, fill=white] at (5.250,-1.500) {\textcolor{black!55}{0.0}};
  
  \node[cell, fill=cmE!0!cmF] at (0.000,-2.250) {\textcolor{white}{100.0}};
  \node[cell, fill=cmE!35!cmF] at (1.050,-2.250) {\textcolor{white}{92.9}};
  \node[cell, fill=cmC!61!cmD] at (2.100,-2.250) {\textcolor{black}{47.9}};
  \node[cell, fill=cmD!13!cmE] at (3.150,-2.250) {\textcolor{white}{77.5}};
  \node[cell, fill=white] at (4.200,-2.250) {\textcolor{black!55}{0.0}};
  \node[cell, fill=cmE!79!cmF] at (5.250,-2.250) {\textcolor{white}{84.3}};
  
  \node[cell, fill=cmE!0!cmF] at (0.000,-3.000) {\textcolor{white}{100.0}};
  \node[cell, fill=cmC!86!cmD] at (1.050,-3.000) {\textcolor{black}{42.9}};
  \node[cell, fill=cmB!4!cmC] at (2.100,-3.000) {\textcolor{black}{39.3}};
  \node[cell, fill=cmE!0!cmF] at (3.150,-3.000) {\textcolor{white}{100.0}};
  \node[cell, fill=cmE!0!cmF] at (4.200,-3.000) {\textcolor{white}{100.0}};
  \node[cell, fill=cmE!44!cmF] at (5.250,-3.000) {\textcolor{white}{91.3}};
  
  \node[cell, fill=cmE!0!cmF] at (0.000,-3.750) {\textcolor{white}{100.0}};
  \node[cell, fill=cmE!0!cmF] at (1.050,-3.750) {\textcolor{white}{100.0}};
  \node[cell, fill=cmE!46!cmF] at (2.100,-3.750) {\textcolor{white}{90.8}};
  \node[cell, fill=cmE!0!cmF] at (3.150,-3.750) {\textcolor{white}{100.0}};
  \node[cell, fill=white] at (4.200,-3.750) {\textcolor{black!55}{0.0}};
  \node[cell, fill=cmE!46!cmF] at (5.250,-3.750) {\textcolor{white}{90.8}};
  
  \node[anchor=north east, rotate=35, inner sep=1pt] at (0.000,-4.175) {\Gemma{}};
  \node[anchor=north east, rotate=35, inner sep=1pt] at (1.050,-4.175) {\LlamaSmall{}};
  \node[anchor=north east, rotate=35, inner sep=1pt] at (2.100,-4.175) {\LlamaLarge{}};
  \node[anchor=north east, rotate=35, inner sep=1pt] at (3.150,-4.175) {\Mistral{}};
  \node[anchor=north east, rotate=35, inner sep=1pt] at (4.200,-4.175) {\QwenMid{}};
  \node[anchor=north east, rotate=35, inner sep=1pt] at (5.250,-4.175) {\QwenLarge{}};
  \node[anchor=east, inner sep=1pt] at (-0.575,0.000) {\Gemma{}};
  \node[anchor=east, inner sep=1pt] at (-0.575,-0.750) {\LlamaSmall{}};
  \node[anchor=east, inner sep=1pt] at (-0.575,-1.500) {\LlamaLarge{}};
  \node[anchor=east, inner sep=1pt] at (-0.575,-2.250) {\Mistral{}};
  \node[anchor=east, inner sep=1pt] at (-0.575,-3.000) {\QwenMid{}};
  \node[anchor=east, inner sep=1pt] at (-0.575,-3.750) {\QwenLarge{}};
  \node[anchor=north, font=\bfseries] at (2.625,-6.125) {};
  \node[anchor=south, font=\bfseries, rotate=90] at (-2.725,-1.875) {Execution Model};

\end{tikzpicture}
        }
        \caption{(\(T=0.0\))}
        \label{fig:crossModel:a}

    \end{subfigure}
    \hfill
    \begin{subfigure}[t]{0.51\linewidth}
        \centering
        \resizebox{\linewidth}{!}{
            \newcommand{\unknown}{?.?}
\begin{tikzpicture}[
    every node/.style={font=\scriptsize},
    cell/.style={draw=black!40, line width=0.3pt, minimum width=1.05cm, minimum height=0.75cm, inner sep=0pt, outer sep=0pt, anchor=center},
  ]
  \definecolor{cmA}{RGB}{255,255,217}
  \definecolor{cmB}{RGB}{199,233,180}
  \definecolor{cmC}{RGB}{127,205,187}
  \definecolor{cmD}{RGB}{65,182,196}
  \definecolor{cmE}{RGB}{44,127,184}
  \definecolor{cmF}{RGB}{37,52,148}
  
  \node[anchor=center, inner sep=1pt] at (0.000,0.750) {\textcolor{red}{68}};
  \node[anchor=center, inner sep=1pt] at (1.050,0.750) {\textcolor{red}{109}};
  \node[anchor=center, inner sep=1pt] at (2.100,0.750) {\textcolor{red}{38}};
  \node[anchor=center, inner sep=1pt] at (3.150,0.750) {\textcolor{red}{93}};
  \node[anchor=center, inner sep=1pt] at (4.200,0.750) {\textcolor{red}{97}};
  \node[anchor=center, inner sep=1pt] at (5.250,0.750) {\textcolor{red}{48}};
  \node[anchor=east, inner sep=1pt] at (-0.575,0.750) {\textcolor{red}{}};
  
  \node[cell, fill=cmE!0!cmF] at (0.000,0.000) {\textcolor{white}{100.0}};
  \node[cell, fill=cmE!35!cmF] at (1.050,0.000) {\textcolor{white}{92.9}};
  \node[cell, fill=cmA!82!cmB] at (2.100,0.000) {\textcolor{black}{3.6}};
  \node[cell, fill=cmE!35!cmF] at (3.150,0.000) {\textcolor{white}{92.9}};
  \node[cell, fill=cmE!0!cmF] at (4.200,0.000) {\textcolor{white}{100.0}};
  \node[cell, fill=cmC!85!cmD] at (5.250,0.000) {\textcolor{black}{42.9}};
  
  \node[cell, fill=cmB!96!cmC] at (0.000,-0.750) {\textcolor{black}{20.8}};
  \node[cell, fill=cmE!18!cmF] at (1.050,-0.750) {\textcolor{white}{96.4}};
  \node[cell, fill=cmC!50!cmD] at (2.100,-0.750) {\textcolor{white}{50.0}};
  \node[cell, fill=cmA!8!cmB] at (3.150,-0.750) {\textcolor{black}{18.4}};
  \node[cell, fill=cmE!18!cmF] at (4.200,-0.750) {\textcolor{white}{96.4}};
  \node[cell, fill=white] at (5.250,-0.750) {\textcolor{black!55}{0.0}};
  
  \node[cell, fill=white] at (0.000,-1.500) {\textcolor{black!55}{0.0}};
  \node[cell, fill=cmE!0!cmF] at (1.050,-1.500) {\textcolor{white}{100.0}};
  \node[cell, fill=cmE!25!cmF] at (2.100,-1.500) {\textcolor{white}{95.0}};
  \node[cell, fill=cmD!0!cmE] at (3.150,-1.500) {\textcolor{white}{80.0}};
  \node[cell, fill=cmE!0!cmF] at (4.200,-1.500) {\textcolor{white}{100.0}};
  \node[cell, fill=cmC!29!cmD] at (5.250,-1.500) {\textcolor{white}{54.2}};
  
  \node[cell, fill=cmE!35!cmF] at (0.000,-2.250) {\textcolor{white}{92.9}};
  \node[cell, fill=cmE!81!cmF] at (1.050,-2.250) {\textcolor{white}{83.7}};
  \node[cell, fill=cmE!35!cmF] at (2.100,-2.250) {\textcolor{white}{92.9}};
  \node[cell, fill=cmE!0!cmF] at (3.150,-2.250) {\textcolor{white}{100.0}};
  \node[cell, fill=cmD!48!cmE] at (4.200,-2.250) {\textcolor{white}{70.4}};
  \node[cell, fill=cmD!48!cmE] at (5.250,-2.250) {\textcolor{white}{70.4}};
  
  \node[cell, fill=cmE!35!cmF] at (0.000,-3.000) {\textcolor{white}{92.9}};
  \node[cell, fill=cmE!0!cmF] at (1.050,-3.000) {\textcolor{white}{100.0}};
  \node[cell, fill=cmD!23!cmE] at (2.100,-3.000) {\textcolor{white}{75.3}};
  \node[cell, fill=cmC!70!cmD] at (3.150,-3.000) {\textcolor{black}{46.0}};
  \node[cell, fill=cmE!40!cmF] at (4.200,-3.000) {\textcolor{white}{91.9}};
  \node[cell, fill=cmE!0!cmF] at (5.250,-3.000) {\textcolor{white}{100.0}};
  
  \node[cell, fill=cmD!25!cmE] at (0.000,-3.750) {\textcolor{white}{75.0}};
  \node[cell, fill=cmE!0!cmF] at (1.050,-3.750) {\textcolor{white}{100.0}};
  \node[cell, fill=cmE!41!cmF] at (2.100,-3.750) {\textcolor{white}{91.7}};
  \node[cell, fill=cmC!8!cmD] at (3.150,-3.750) {\textcolor{white}{58.3}};
  \node[cell, fill=cmE!0!cmF] at (4.200,-3.750) {\textcolor{white}{100.0}};
  \node[cell, fill=cmE!0!cmF] at (5.250,-3.750) {\textcolor{white}{100.0}};
  
  \node[anchor=north east, rotate=35, inner sep=1pt] at (0.000,-4.175) {\Gemma{}};
  \node[anchor=north east, rotate=35, inner sep=1pt] at (1.050,-4.175) {\LlamaSmall{}};
  \node[anchor=north east, rotate=35, inner sep=1pt] at (2.100,-4.175) {\LlamaLarge{}};
  \node[anchor=north east, rotate=35, inner sep=1pt] at (3.150,-4.175) {\Mistral{}};
  \node[anchor=north east, rotate=35, inner sep=1pt] at (4.200,-4.175) {\QwenMid{}};
  \node[anchor=north east, rotate=35, inner sep=1pt] at (5.250,-4.175) {\QwenLarge{}};
  
  \node[anchor=east, inner sep=1pt] at (-0.575,0.000) {};
  \node[anchor=east, inner sep=1pt] at (-0.575,-0.750) {};
  \node[anchor=east, inner sep=1pt] at (-0.575,-1.500) {};
  \node[anchor=east, inner sep=1pt] at (-0.575,-2.250) {};
  \node[anchor=east, inner sep=1pt] at (-0.575,-3.000) {};
  \node[anchor=east, inner sep=1pt] at (-0.575,-3.750) {};
  
  \node[anchor=north, font=\small\bfseries] at (2.625,-6.125) {};
  \node[anchor=south, font=\small\bfseries, rotate=90] at (-2.725,-1.875) {};
  
  \node[anchor=center, font=\small\bfseries] at (2.625, 1.35) {};
  
  \fill[cmA!95!cmB] (6.375,-4.125) rectangle (6.725,-4.031);
  \fill[cmA!85!cmB] (6.375,-4.031) rectangle (6.725,-3.938);
  \fill[cmA!75!cmB] (6.375,-3.938) rectangle (6.725,-3.844);
  \fill[cmA!65!cmB] (6.375,-3.844) rectangle (6.725,-3.750);
  \fill[cmA!55!cmB] (6.375,-3.750) rectangle (6.725,-3.656);
  \fill[cmA!45!cmB] (6.375,-3.656) rectangle (6.725,-3.562);
  \fill[cmA!35!cmB] (6.375,-3.562) rectangle (6.725,-3.469);
  \fill[cmA!25!cmB] (6.375,-3.469) rectangle (6.725,-3.375);
  \fill[cmA!15!cmB] (6.375,-3.375) rectangle (6.725,-3.281);
  \fill[cmA!5!cmB]  (6.375,-3.281) rectangle (6.725,-3.188);
  
  \fill[cmB!95!cmC] (6.375,-3.188) rectangle (6.725,-3.094);
  \fill[cmB!85!cmC] (6.375,-3.094) rectangle (6.725,-3.000);
  \fill[cmB!75!cmC] (6.375,-3.000) rectangle (6.725,-2.906);
  \fill[cmB!65!cmC] (6.375,-2.906) rectangle (6.725,-2.812);
  \fill[cmB!55!cmC] (6.375,-2.812) rectangle (6.725,-2.719);
  \fill[cmB!45!cmC] (6.375,-2.719) rectangle (6.725,-2.625);
  \fill[cmB!35!cmC] (6.375,-2.625) rectangle (6.725,-2.531);
  \fill[cmB!25!cmC] (6.375,-2.531) rectangle (6.725,-2.438);
  \fill[cmB!15!cmC] (6.375,-2.438) rectangle (6.725,-2.344);
  \fill[cmB!5!cmC]  (6.375,-2.344) rectangle (6.725,-2.250);
  
  \fill[cmC!95!cmD] (6.375,-2.250) rectangle (6.725,-2.156);
  \fill[cmC!85!cmD] (6.375,-2.156) rectangle (6.725,-2.062);
  \fill[cmC!75!cmD] (6.375,-2.062) rectangle (6.725,-1.969);
  \fill[cmC!65!cmD] (6.375,-1.969) rectangle (6.725,-1.875);
  \fill[cmC!55!cmD] (6.375,-1.875) rectangle (6.725,-1.781);
  \fill[cmC!45!cmD] (6.375,-1.781) rectangle (6.725,-1.688);
  \fill[cmC!35!cmD] (6.375,-1.688) rectangle (6.725,-1.594);
  \fill[cmC!25!cmD] (6.375,-1.594) rectangle (6.725,-1.500);
  \fill[cmC!15!cmD] (6.375,-1.500) rectangle (6.725,-1.406);
  \fill[cmC!5!cmD]  (6.375,-1.406) rectangle (6.725,-1.312);
  
  \fill[cmD!95!cmE] (6.375,-1.312) rectangle (6.725,-1.219);
  \fill[cmD!85!cmE] (6.375,-1.219) rectangle (6.725,-1.125);
  \fill[cmD!75!cmE] (6.375,-1.125) rectangle (6.725,-1.031);
  \fill[cmD!65!cmE] (6.375,-1.031) rectangle (6.725,-0.938);
  \fill[cmD!55!cmE] (6.375,-0.938) rectangle (6.725,-0.844);
  \fill[cmD!45!cmE] (6.375,-0.844) rectangle (6.725,-0.750);
  \fill[cmD!35!cmE] (6.375,-0.750) rectangle (6.725,-0.656);
  \fill[cmD!25!cmE] (6.375,-0.656) rectangle (6.725,-0.562);
  \fill[cmD!15!cmE] (6.375,-0.562) rectangle (6.725,-0.469);
  \fill[cmD!5!cmE]  (6.375,-0.469) rectangle (6.725,-0.375);
  
  \fill[cmE!95!cmF] (6.375,-0.375) rectangle (6.725,-0.281);
  \fill[cmE!85!cmF] (6.375,-0.281) rectangle (6.725,-0.188);
  \fill[cmE!75!cmF] (6.375,-0.188) rectangle (6.725,-0.094);
  \fill[cmE!65!cmF] (6.375,-0.094) rectangle (6.725,0.000);
  \fill[cmE!55!cmF] (6.375,0.000)  rectangle (6.725,0.094);
  \fill[cmE!45!cmF] (6.375,0.094)  rectangle (6.725,0.188);
  \fill[cmE!35!cmF] (6.375,0.188)  rectangle (6.725,0.281);
  \fill[cmE!25!cmF] (6.375,0.281)  rectangle (6.725,0.375);
  \fill[cmE!15!cmF] (6.375,0.375)  rectangle (6.725,0.469);
  \fill[cmE!5!cmF]  (6.375,0.469)  rectangle (6.725,0.563);
  
  \draw[black!50, line width=0.3pt] (6.375,-4.125) rectangle (6.725,0.563);
  \node[anchor=west, inner sep=1pt] at (6.775,-4.125) {0};
  \node[anchor=west, inner sep=1pt] at (6.775,-3.188) {20};
  \node[anchor=west, inner sep=1pt] at (6.775,-2.250) {40};
  \node[anchor=west, inner sep=1pt] at (6.775,-1.312) {60};
  \node[anchor=west, inner sep=1pt] at (6.775,-0.375) {80};
  \node[anchor=west, inner sep=1pt] at (6.775,0.563)  {100};
\end{tikzpicture}
        }
        
        \caption{(\(T=0.5\))}
        \label{fig:crossModel:b}
    \end{subfigure}
        \vspace{0.5em}
    \centering
    \textbf{Source Model (Minimized Prompt Generator)}
    \caption{Cross-Model Transfer Accuracy (\% Constraint Fidelity).
    Rows index the execution model; columns index the source model
    that produced the minimized prompt.
    Diagonal cells show each model's 10-run self-accuracy
    under its own minimized prompt;
    off-diagonal cells show transfer accuracy
    when the minimized prompt from the column model
    is executed by the row model.
    Empty (zero) cells indicate that no valid output was produced.
    The red numbers above each column show the minimized prompt length (in tokens) produced by that source model.}
    \label{fig:crossModel}
\end{figure*}

To answer RQ4,
we cross-execute every minimized prompt on every evaluated model.
For each ordered pair \((M_s, M_e)\)---source model \(M_s\)
that produced the minimized prompt,
execution model \(M_e\) that consumes it---we
ran 10 independent generations
and measured constraint fidelity against the full-prompt baseline.
\Cref{fig:crossModel} reports the resulting \(6{\times}6\) transfer matrices for \(T=0.0\) and \(T=0.5\);
diagonal cells repeat the single-model self-accuracy from RQ3,
while off-diagonal cells isolate the transfer effect.

\begin{figure}[tp]
    \centering
    \newcommand{\IDwidth}{0.72\columnwidth}%
    \newcommand{\IDheight}{0.55\columnwidth}%
    \providecommand{\IDwidth}{0.78\columnwidth}
\providecommand{\IDheight}{0.50\columnwidth}
\begin{tikzpicture}
\definecolor{colorA}{RGB}{100,140,190}   
\definecolor{colorB}{RGB}{220,150,60}    
\definecolor{darktext}{RGB}{43,43,43}

\begin{axis}[
    width=\IDwidth,
    height=\IDheight,
    scale only axis,
    ybar,
    bar width=5pt,
    ymin=0, ymax=165,
    ytick={0,20,40,60,80,100,120,140,160},
    ylabel={\textbf{Example Length (Tokens)}},
    ylabel style={text=colorA, font=\scriptsize},
    y tick label style={text=colorA, font=\scriptsize},
    xlabel={\textbf{Source Model}},
    xlabel style={text=darktext, font=\scriptsize},
    x tick label style={
        text=darktext,
        font=\scriptsize,
        rotate=35,
        anchor=north east,
        inner sep=1pt,
        xshift=2pt
    },
    symbolic x coords={llama_3.3_70b,qwen_72b,qwen_32b,gemma_2_9b_it,llama_3.1_8b,mistral_24b},
    xtick=data,
    enlarge x limits=0.12,
    title={\textbf{Information Density vs. Cross-Model Robustness}},
    title style={text=darktext, font=\footnotesize, yshift=1ex},
    axis line style={thick},
    tick style={thick, black},
    ytick pos=left,
    xtick pos=bottom,
    ytick align=outside,
    xtick align=outside,
    legend style={
        at={(0.5,-0.48)},
        anchor=north,
        legend columns=2,
        font=\scriptsize,
        draw=none,
        fill=none,
        column sep=1.5em,
    },
    legend image code/.code={
        \draw[#1, draw=none] (0cm,-0.1cm) rectangle (0.3cm,0.15cm);
        \node[mark size=0pt, #1] at (0.0cm,0.00cm) {\pgfuseplotmark{*}};
    },
]

\addplot[
    fill=colorA,
    draw=none,
    bar shift=-3pt,
] coordinates {
    (llama_3.3_70b,29)
    (qwen_72b,49)
    (qwen_32b,63)
    (gemma_2_9b_it,90)
    (llama_3.1_8b,114)
    (mistral_24b,154)
};
\addlegendentry{\textcolor{colorA}{$T{=}0.0$}}

\addplot[
    fill=colorB,
    draw=none,
    bar shift=3pt,
] coordinates {
    (llama_3.3_70b,38)
    (qwen_72b,48)
    (qwen_32b,97)
    (gemma_2_9b_it,68)
    (llama_3.1_8b,109)
    (mistral_24b,93)
};
\addlegendentry{\textcolor{colorB}{$T{=}0.5$}}

\end{axis}

\begin{axis}[
    width=\IDwidth,
    height=\IDheight,
    scale only axis,
    ymin=0, ymax=132,
    ytick={0,20,40,60,80,100,120},
    ylabel={\textbf{Universal Transfer Accuracy (\%)}},
    ylabel style={text=darktext, font=\scriptsize},
    y tick label style={text=darktext, font=\scriptsize},
    symbolic x coords={llama_3.3_70b,qwen_72b,qwen_32b,gemma_2_9b_it,llama_3.1_8b,mistral_24b},
    xtick=data,
    enlarge x limits=0.12,
    axis x line=none,
    axis y line*=right,
    axis line style={thick},
    tick style={thick, black},
    ytick align=outside,
]

\addplot[
    color=black,
    dotted,
    line width=1.0pt,
    forget plot,
    mark=none
] coordinates {
    (llama_3.3_70b,100)
    (mistral_24b,100)
};

\addplot[
    color=colorA,
    dashed,
    line width=1.5pt,
    mark=*,
    mark options={solid, scale=1.5, fill=colorA},
    forget plot,
] coordinates {
    (llama_3.3_70b,43.1)
    (qwen_72b,49.4)
    (qwen_32b,0)
    (gemma_2_9b_it,97.9)
    (llama_3.1_8b,65.7)
    (mistral_24b,89.8)
};

\addplot[
    color=colorB,
    dashed,
    line width=1.5pt,
    mark=square*,
    mark options={solid, scale=1.5, fill=colorB},
    forget plot,
] coordinates {
    (llama_3.3_70b,62.7)
    (qwen_72b,53.5)
    (qwen_32b,93.4)
    (gemma_2_9b_it,56.3)
    (llama_3.1_8b,95.3)
    (mistral_24b,59.1)
};

\end{axis}
\end{tikzpicture}%
    \caption{Information density versus cross-model robustness. For each
source model (x-axis, ordered by example length), blue bars show the
length of its minimized exemplar in tokens (left axis), and the red
dashed line shows the mean transfer accuracy of that prompt when
executed by the other five models (right axis).}
    \label{fig:infoDen}
\end{figure}
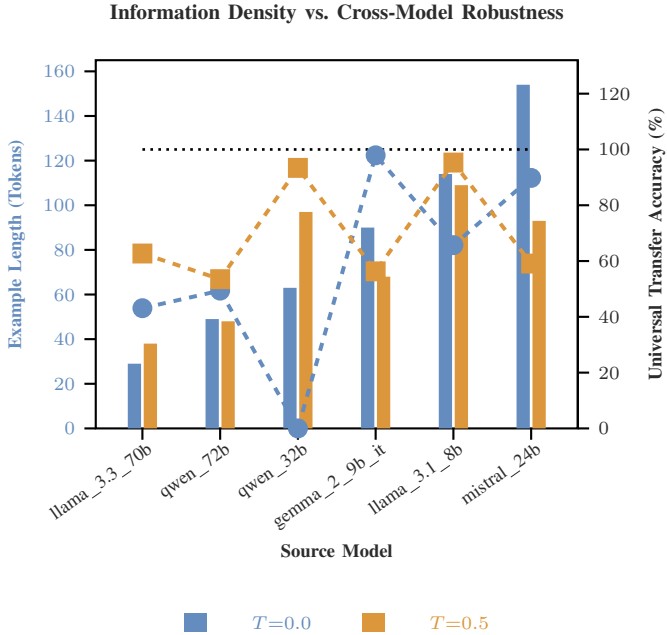

\noindent\textbf{Transfer is the common case, not the exception.}
At \(T=0.0\), of the 30 off-diagonal pairs,
22 yield a non-empty output that parses into the expected
propositional form,
and the average accuracy of those 22 successful transfers is 78.1\%.
Eight of the 30 off-diagonal pairs meet
100\% constraint fidelity. That is, they reproduce the full-prompt baseline
as exactly as the source model does on itself, and
thirteen reach at least 90\%.
The minimized core is therefore not a model-specific artifact.
In the majority of cases
it captures content that is sufficient for a different model
to reconstruct the same propositional output.
At \(T=0.5\) (\Cref{fig:crossModel:b}), 28 of the 30 off-diagonal pairs yield non-empty output
that parses into the expected propositional form,
with a mean accuracy of 75.1\%; seven of those 28 reach 100\% fidelity.

\noindent\textbf{Source models differ sharply in how portable their prompts are.}
Reading \Cref{fig:crossModel:a} by column isolates how well each source model's minimized prompt travels.
Two source models produce universally portable prompts at \(T=0.0\):
every other execution model was able to use
the minimized prompts from \Gemma{}
(mean 96.7\% across five targets, all five non-empty)
and from \Mistral{}
(mean 88.8\%, all five non-empty).
Turning to \Cref{fig:crossModel:b}, at \(T=0.5\) the prompts from \LlamaSmall{} and \QwenMid{}
were successfully consumed by all models,
with mean accuracies of 95.3\% and 93.4\% respectively.
At the opposite extreme,
the \QwenMid{} prompt at \(T=0.0\) fails to transfer to any other model
(0/5 non-empty transfers).
\LlamaSmall{}, \LlamaLarge{}, and \QwenLarge{} sit in between,
with four of five targets producing valid output at \(T=0.0\).
At \(T=0.5\), \LlamaSmall{}, \LlamaLarge{}, \Mistral{}, and \QwenMid{} successfully transferred to all models.
Source portability is therefore not monotone in model size:
the largest source model in the study (\QwenLarge{})
is less portable than a 9B source (\Gemma{}),
and the single worst source (\QwenMid{})
sits in the middle of the size range.

\noindent\textbf{Execution models differ in what they can consume.}
Reading the matrix in \Cref{fig:crossModel:a} by row isolates the execution side.
\Gemma{}, \Mistral{}, and \QwenLarge{}
each successfully consume prompts from four of the other five sources at \(T=0.0\);
at \(T=0.5\) (\Cref{fig:crossModel:b}) they consume prompts from all sources, as does \QwenMid{}.
The two Llama variants consume two and three of five sources respectively at \(T=0.0\),
but reach four of five at \(T=0.5\).
\QwenMid{} was able to consume all five in both temperature settings.
The high execution-side scores of \QwenLarge{}
(mean non-empty transfer of 95.4\%)
indicate that these models tolerate a wide variety of minimization styles,
even when their own minimized prompt is less universally portable.

\noindent\textbf{Information density tracks transfer robustness.}
\Cref{fig:infoDen} overlays,
for each source model,
the length in tokens of its minimized exemplar (left axis,  blue and orange bars)
against the mean transfer accuracy of that prompt
when executed by the other five models (right axis, dashed lines).
Very short minimization fails to carry enough signal for other models:
\LlamaLarge{}, which produced the shortest minimized exemplar
(\(\approx 29\) tokens),
reaches only about 49\% at \(T=0.0\) and 62.7\% at \(T=0.5\) mean transfer accuracy,
and \QwenMid{}, whose minimization is similarly aggressive at \(T=0.0\),
transfers to no other model at all. \QwenMid{}, however, transfers to all models
at 93.4\% mean accuracy at \(T=0.5\), in which case its minimized prompt
is considerably longer (\(\approx 97\) tokens).
In contrast, \Gemma{} (\(\approx 90\) tokens at \(T=0.0\)) and \Mistral{}
(\(\approx 153\) tokens at \(T=0.0\)) achieve universal transfer at or above 88.0\% mean accuracy.
When temperature is raised to 0.5, these models compress their prompts further
(to 68 and 93~tokens respectively), and their universal transferability decreases accordingly.
Longer minimized exemplars transfer more reliably:
\Gemma{} (\(\approx 90\) tokens) and \Mistral{} (\(\approx 153\) tokens)
achieve universal transfer at or above 90\% mean accuracy.
In other words,
a source model that compresses too aggressively
overfits to its own priors
and strips content that other models still need,
whereas a source that leaves more structural scaffolding in place
produces a more model-agnostic minimal core.

\noindent\textbf{Transferred prompts can outperform self-accuracy.}
Eight off-diagonal cells reached 100\%;
notably, \Mistral{}’s minimized prompt was successfully executed
by \LlamaLarge{}, \QwenLarge{}, and \Gemma{} at 100.0\%.
Some models,
conditioned on a different model's minimization,
reproduce the baseline constraints more reliably
than when conditioned on its own minimization
(e.g.\ \QwenLarge{} achieves 90.8\% self-accuracy at \(T=0.0\)
but executes \Gemma{}, \LlamaSmall{}, and \Mistral{} prompts at 100.0\%).
This phenomenon persists across sampling temperatures;
at \(T=0.5\), \LlamaSmall{} achieves 96.4\% self-accuracy under its own prompt,
yet that same minimized prompt serves as a perfect instructional scaffold
(100.0\% execution accuracy) for \QwenMid{}, \QwenLarge{}, and \LlamaLarge{}.
This is consistent with the reading that minimization is not
strictly a per-model operation:
a prompt minimized by one model can serve as a better scaffold for a second model
than the second model's own minimization.

\begin{result}
\textbf{RQ4.}
The minimized prompt transfers across models in (22 of 30 at T=0.0; 28 of 30 at T=0.5) off-diagonal pairs.
Transfer is not determined by model size:
aggressive minimization (fewer tokens) overfits the prompt to the source model's priors, while prompts that retain more structural scaffolding (more tokens) serve as highly portable instructions that occasionally outperform a target model's own self-accuracy.
\end{result}

\section{Discussion}\label{sec:discussion}

\subsection{Model Scale as a Proxy for Scaffold Independence}

We model achievable compression as a function of model size~$S$
(in billions of parameters) and sampling temperature~$T$,
using ordinary least squares across all 12 model/temperature runs ($n = 12$).
The saturated model is
\[
\mu\{R \mid S, T\} = \beta_0 + \beta_1 S + \beta_2 T + \beta_3 (S \times T)
\]
and is reported in \Cref{tab:regModel1}.

\begin{table}[tp]
\centering
\caption{Saturated model: $R \sim S + T + S{\times}T$.
         $R^2 = 0.566$, Adj.\ $R^2 = 0.403$.}
\begin{tabular}{lrrrr}
\toprule
  & Estimate & Std.\ Error & $t$ & $\Pr(>|t|)$ \\
\midrule
(Intercept) & 46.41 & 8.47 & 5.48 & $< 0.001$ \\
$S$ & 0.51 & 0.19 & 2.67 & 0.029 \\
$T$ & 14.40 & 23.95 & 0.60 & 0.564 \\
$S{\times}T$ & $-0.32$ & 0.54 & $-0.60$ & 0.564 \\
\bottomrule
\end{tabular}

\label{tab:regModel1}
\end{table}

VIF values for $S$, $T$, and $S \times T$ are 2.00, 2.87, and 3.87,
indicating no problematic multicollinearity.
The interaction term $S \times T$ has the highest $p$-value (0.564)
and is removed first, giving \Cref{tab:regModel2}.

\begin{table}[tp]
\centering
\caption{Reduced model: $R \sim S + T$.
         $R^2 = 0.546$, Adj.\ $R^2 = 0.446$.}
\begin{tabular}{lrrrr}
\toprule
  & Estimate & Std.\ Error & $t$ & $\Pr(>|t|)$ \\
\midrule
(Intercept) & 49.31 & 6.70 & 7.36 & $< 0.001$ \\
$S$ & 0.43 & 0.13 & 3.29 & 0.009 \\
$T$ & 2.77 & 13.63 & 0.20 & 0.844 \\
\bottomrule
\end{tabular}

\label{tab:regModel2}
\end{table}

Temperature remains far from significance ($p = 0.844$).
An ANOVA comparison confirms that adding $T$ to a size-only model
contributes no explanatory power ($F = 0.04$, $p = 0.844$).
Removing $T$ yields the final model in \Cref{tab:regModel3}.

\begin{table}[tp]
\centering
\caption{Final model: $R \sim S$.\quad
         $R^2 = 0.544$, Adj.\ $R^2 = 0.499$.}
\begin{tabular}{lrrrr}
\toprule
  & Estimate & Std.\ Error & $t$ & $\Pr(>|t|)$ \\
\midrule
(Intercept) & 50.00 & 5.49 & 9.11 & $< 0.001$ \\
$S$ & 0.43 & 0.12 & 3.46 & 0.006 \\
\bottomrule
\end{tabular}

\label{tab:regModel3}
\end{table}

The final equation is
$\mu\{R \mid S\} = 0.43 \cdot S + 50.0$,
shown as the fitted line in \Cref{fig:regressionPlot}.
Each additional 10B parameters corresponds to 4.3 percentage points
of additional achievable compression.
There is no significant effect of sampling temperature on compression rate
in any model formulation tested.

However, the size-only model explains just over half the variance
($R^2 = 0.544$), and conflates parameter count with architecture,
training data, and alignment strategy.
As we show below, a categorical \emph{family} variable
captures substantially more variance ($R^2 = 0.912$),
indicating that raw parameter count is a useful but crude first-order proxy
for scaffold independence.

\noindent\textbf{Log-scale comparison.}
Because model parameter counts often span orders of magnitude,
we also fit the log-linear family $R \sim \log S + T + \log S \times T$
and reduce it by the same stepwise procedure.
Neither $T$ ($p = 0.867$) nor $\log S \times T$ ($p = 0.565$) are significant,
yielding $R \sim \log S$ as the log-scale final model.
\Cref{tab:regAIC} compares the two final single-predictor models directly via AIC.
The linear model is preferred by $\Delta\mathrm{AIC} = 3.83$,
confirming that the relationship between capacity and compression
is better described as linear than as logarithmic
across the 8B--72B range studied.

\begin{table}[tp]
\centering
\caption{AIC comparison of final single-predictor models ($n = 12$).
         Lower AIC indicates a better fit penalized for model complexity.}
\begin{tabular}{lrrrr}
\toprule
Model & $R^2$ & Adj.\ $R^2$ & AIC & $\Delta$AIC \\
\midrule
$R \sim S$ & 0.544 & 0.499 & 93.90 & 0.00 \\
$R \sim \log S$ & 0.373 & 0.310 & 97.73 & $+3.83$ \\
\bottomrule
\end{tabular}

\label{tab:regAIC}
\end{table}

\noindent\textbf{Architecture family.}
The large residuals from the linear model---\Gemma{} exceeds
its size-based prediction by 14.3 points, \Mistral{} falls short by 16.2---suggest
that model architecture and training data contribute to scaffold independence
independently of parameter count.
To quantify this, we replace $S$ with a categorical \emph{family} variable $F$
(six levels; \LlamaSmall{} as reference),
resulting in the saturated family model $R \sim F + T + F \times T$.
Following the same reduction procedure, $F \times T$ is not significant
($F = 0.29$, $p = 0.883$ for the joint interaction test),
and $T$ is again not significant after removing the interaction ($p = 0.743$).
The reduced family model $R \sim F$ is reported in \Cref{tab:regFamily}.

\begin{table}[tp]
\centering
\caption{Family model: $R \sim F$ (\LlamaSmall{} as reference).
         $R^2 = 0.912$, Adj.\ $R^2 = 0.838$.}
\begin{tabular}{lrrrr}
\toprule
  & Estimate & Std.\ Error & $t$ & $\Pr(>|t|)$ \\
\midrule
(I) [\LlamaSmall{}] & 51.70 & 4.51 & 11.47 & $< 0.001$ \\
\Gemma{} & $+16.45$ & 6.37 & 2.58 & 0.042 \\
\Mistral{} & $-7.65$ & 6.37 & $-1.20$ & 0.275 \\
\QwenMid{} & $+11.35$ & 6.37 & 1.78 & 0.125 \\
\LlamaLarge{} & $+34.10$ & 6.37 & 5.35 & 0.002 \\
\QwenLarge{} & $+27.40$ & 6.37 & 4.30 & 0.005 \\
\bottomrule
\end{tabular}

\label{tab:regFamily}
\end{table}

\Cref{tab:regSummary} summarizes all the models.
The family model improves fit dramatically over the size model
($\Delta\mathrm{AIC} = -11.71$, $R^2 = 0.912$ vs.\ $0.544$),
demonstrating that architecture and training choices explain a large share
of the variance that the raw parameter count cannot capture.

\begin{table}[tp]
\centering
\caption{Summary of all regression models ($n = 12$).}
\begin{tabular}{lrrrr}
\toprule
Model & $R^2$ & Adj.\ $R^2$ & AIC & $\Delta$AIC \\
\midrule
$R \sim S$ & 0.544 & 0.499 & 93.90 & 0.00 \\
$R \sim \log S$ & 0.373 & 0.310 & 97.73 & $+3.83$ \\
$R \sim S + T$ & 0.546 & 0.446 & 95.85 & $+1.95$ \\
$R \sim \log S + T$ & 0.375 & 0.236 & 99.69 & $+5.79$ \\
$R \sim F$ & 0.912 & 0.838 & 82.19 & $-11.71$ \\
$R \sim F + T$ & 0.914 & 0.811 & 83.90 & $-10.00$ \\
\bottomrule
\end{tabular}

\label{tab:regSummary}
\end{table}

Within the family model,
two contrasts are particularly informative.
First, \Gemma{} achieves 16.5 percentage points more compression
than \LlamaSmall{} despite having essentially the same parameter count (9B vs.\ 8B;
$p = 0.042$).
This rules out size as the sole explanation and implicates architectural or
training-data differences between the two families.
Second, \Mistral{} performs 7.7 points below the reference 8B model
(not significant, $p = 0.275$), despite being three times larger.
The Mistral family therefore does not gain the scaffold independence
that its size would predict under the linear model.
The two large models---\LlamaLarge{} and \QwenLarge{}---both show strongly
significant positive contrasts ($p = 0.002$ and $p = 0.005$),
consistent with the size effect identified earlier,
but their magnitude differs (34.1 vs.\ 27.4 points above reference),
again pointing to family-level variation beyond size.

\subsection{Universal Encoders and Universal Decoders}

The cross-model transfer matrix (\Cref{fig:crossModel})
exposes an asymmetry that a single-model view of minimization hides:
source portability and execution robustness are two distinct model properties,
and a model can be strong on one axis and weak on the other.

\noindent\textbf{Universal encoders.}
Inspecting \Cref{fig:crossModel} columnwise,
two source models produce minimized prompts
that every other model evaluated can successfully consume:
\Gemma{} (mean transfer accuracy 96.7\% across five targets)
and \Mistral{} (88.8\%, all five targets non-empty).
Neither is the largest model in the study.
Their minimized prompts are comparatively long
(roughly 90 and 154 tokens respectively, see \Cref{fig:infoDen}),
indicating that these models leave enough structural scaffolding in place
to serve as a model-agnostic minimal core.
We call such models \emph{universal encoders}.

\noindent\textbf{Universal decoders.}
Inspecting \Cref{fig:crossModel} rowwise,
some models reliably consume minimized prompts produced by other sources
regardless of how aggressively those sources compress.
\QwenLarge{} is the clearest example,
reaching a mean non-empty transfer accuracy of 95.4\%
on prompts from four of the five other source models,
notably including \Gemma{}, \LlamaSmall{}, and \Mistral{} prompts, on which it scores 100\%,
despite achieving only 90.8\% self-accuracy on its own minimized prompt.
\QwenMid{} is a second example, which shows a high ability to execute on all sources (74.7\% mean transfer, five of five sources).
We call such models \emph{universal decoders}.

\noindent\textbf{Interaction with compression aggressiveness.}
The \QwenMid{} row is informative:
its own minimized prompt is aggressive enough
that no other model can consume it (0/5 non-empty transfers),
yet \QwenMid{} itself consumes other models' minimizations
on all five targets.
Aggressive self-compression and poor portability therefore co-occur,
consistent with the interpretation that over-compression
encodes model-specific shortcuts.
\Cref{fig:infoDen} makes this quantitative:
the two shortest minimized exemplars (\LlamaLarge{} at $\approx$29 tokens,
\QwenMid{} at $\approx$63 tokens)
correspond to the two worst mean transfer accuracies (43\% and 0\%),
while the two longest exemplars (\Gemma{} and \Mistral{})
correspond to the two universal encoders.

This asymmetry has a practical consequence.
If a minimized prompt must be deployed across heterogeneous models,
a common case in production pipelines with model-swap or fallback logic,
the minimization should be produced on a universal-encoders model
rather than on the aggressively compressing ones,
even when the latter achieves a higher self-compression rate.

\subsection{Implications for Prompt Engineering}
Our findings have three practical implications for prompt engineering.

\begin{enumerate}
\item \textbf{Cross-prompt annotations are redundant.}
The high removal rate of \nonterm{LRel} (72.6\%) is perhaps the most actionable finding for
prompt designers.
The {\small \texttt{LogicalRelationshipWithPreviousPrompt}} block was included in the P2P exemplar
to help the model understand how consecutive prompts relate to one another---refinements,
additions, and shared conjuncts across examples.
Its near-complete dispensability indicates that models process each exemplar
largely independently rather than chaining them relationally.
Prompt designers can therefore omit cross-prompt relationship annotations
without loss of output fidelity,
reducing prompt length and engineering overhead simultaneously.

\item \textbf{Logical identifiers are the causal core.}
The survival of \nonterm{PID}, \nonterm{CID}, \nonterm{Constraint}, and \nonterm{LExp}
suggests that models attend to the \emph{structural pattern}---the
typed slot names, the identifier tokens, and the logical connectives---rather
than to the relational scaffolding that surrounds them.
This aligns with prior work showing that labels in few-shot examples
can be randomly permuted with only modest accuracy loss~\cite{min2022rethinking},
and extends that observation to a finer-grained, grammar-level analysis:
it is the \emph{schema} of the exemplar, not its \emph{content}, that is causally necessary.

\item \textbf{Minimize on a universal encoder, deploy on a universal decoder.}
When a single prompt must serve multiple downstream models,
the minimization source matters more than the minimization rate.
A prompt minimized by \Gemma{} or \Mistral{}
transfers universally,
whereas an equally or more compressed prompt from \QwenMid{}
fails to transfer to any other model.
Pairing a universal-encoder source with a universal-decoder target
(e.g.\ \Mistral{} $\to$ \QwenLarge{})
yields fidelity that can exceed the target's self-accuracy.
\end{enumerate}

\section{Related Work}\label{sec:related}
\subsection{Delta Debugging and Prompt Minimization}
The core of our \framework pipeline is delta debugging, a technique for automatically isolating the minimal cause of a failure. Zeller and Hildebrandt~\cite{zeller2002simplifying} introduced the minimizing delta debugging algorithm (\ddmin) to find the minimal cause of a bug in a failing program. This technique was further improved by Zeller~\cite{zeller2002isolating} so that \ddmin can isolate cause-effect chains in program execution. Furthermore, exploiting structural grammar was an improvement introduced by Misherghi and Su~\cite{misherghi2006hdd} through Hierarchical Delta Debugging.

Beyond Delta Debugging, several techniques were introduced for prompt minimization. Su et al.~\cite{su2022selective} introduced Selective Annotations, which select the most informative examples. This technique led to effectively minimizing the number of examples without losing performance. Additionally, RLPrompt is a reinforcement-learning technique introduced by Deng et al.~\cite{deng2022rlprompt} that identifies the most effective combinations of tokens to optimize and simplify prompts for a given task. Shi et al.~\cite{shi2024prompt} introduced Prompt Space, a mathematically grounded framework that selects few-shot exemplars. Prompt Space showed improved performance compared to heuristic prompting strategies such as zero-shot and few-shot.
\subsection{Chain-of-Thought and Few-shot Learning}
Brown et al.~\cite{brown2020} established few-shot prompting as the standard approach to in-context learning. Few-shot prompting allows models to complete tasks by learning from a number of examples included in the prompt. Chain-of-Thought (CoT) prompting emerged as a major development in this field. Wei et al.~\cite{wei2022} showed that when the model is asked to show its intermediate reasoning steps, it can achieve much higher performance on difficult math and logic problems. Because LLMs can be unpredictable (stochastic), Wang et al.~\cite{wang2022self} proposed Self-Consistency to handle such cases. Self-Consistency demonstrates that even if the reasoning process varies, the final logical result should stay consistent across different attempts.

Although long CoT prompts are effective, recent studies have shown that models might not actually need the entire CoT text. Kojima et al.~\cite{kojima2022large} showed that LLMs can act as Zero-Shot Reasoners when asked to use a simple phrase like ``Let's think step by step.'' This suggests that a large portion of the CoT text in few-shot examples may be redundant or unnecessary. Additionally, Reynolds and McDonell~\cite{reynolds2021prompt} argued that few-shot examples serve as ``locative'' signals that guide the model to the correct task environment, rather than teaching it new information. These studies provide the theoretical foundation for our \framework framework. This suggests that within every complex CoT prompt there may be a ``minimal causal core'' that can be extracted without damaging the model's logical performance.
\subsection{Prompt Sensitivity and Reliability}
The non-determinism of LLMs has been highlighted in recent research. This phenomenon creates challenges for downstream tasks. Although deterministic-behavior settings such as setting temperature to zero exist, the model can still produce different outputs. Astekin et al.~\cite{astekin2024exploratory} showed that LLMs produced inconsistent log-parsing templates across multiple runs, which raises serious concerns about reproducibility in software engineering. Similarly, Atil et al.~\cite{atil2024non} conducted a systematic study to quantify output variability across different benchmarks. Regardless of the deterministic configurations, they found accuracy variation of up to 15\%.
Prompts are highly sensitive to small changes in their structure. Zhao et al.~\cite{zhao2021calibrate} showed that the order and formatting of examples can play a large role in accuracy. A model may therefore produce the right answer using one example but fail if one important word is removed. This instability can lead to security risks. Pearce et al.~\cite{pearce2022asleep} found that a model can produce insecure code because of prompt variations. Moreover, Lu et al.~\cite{lu2022fantastically} found that prompt structure often matters more than the semantic meaning of the text. These findings motivate the need for an approach that tells researchers which specific part of a prompt is important to maintain stable output.

\section{Threats to Validity}\label{sec:threats}
\subsection{External Validity}
Our experiments cover six models and two temperature settings.
While this range spans 8B to 72B parameters
and includes four distinct model families
(Llama, Gemma, Qwen, and Mistral),
the results may not generalize to models trained on different corpora,
to proprietary frontier models,
or to models using alternative decoding strategies
(for example, nucleus or beam search).
The evaluation dataset itself is also narrow:
two inquiries of two prompts each,
drawn from a programming-tasks domain.
Findings about which structural layers are preserved
may shift for prompts in natural-language, mathematical,
or multimodal domains,
and we mitigate this only partially
by releasing the \framework pipeline
so other researchers can run the same analysis
on their own datasets and models.

\subsection{Internal Validity}
The P2P formalization layer introduces a potential source of error.
If the formalization itself is imperfect,
the comparison between baseline and minimized outputs will be unreliable.
We validated P2P on a separate dataset~\cite{alfageeh2025from}
and found it to be highly accurate.
A second internal threat is the non-determinism of the models themselves:
a configuration that matches the baseline on one run
may not match it on re-evaluation.
To mitigate this, \framework classifies any such non-reproducible
configuration as \texttt{UNRESOLVED} rather than \texttt{PRESERVED}
(see \Cref{subsec:dd-verdicts}),
so flaky runs do not cause over-reduction,
although they may cause under-reduction.

\subsection{Construct Validity}
``Propositional output fidelity''
is a proxy for ``task correctness.''
We have shown that minimized prompts preserve propositional output,
but it is possible that they fail to capture some subtle aspects of task intent
that are not reflected in the constraint set.
Our RQ3 accuracy experiments provide partial evidence
that the proxy is reasonable---%
four of six models reproduce full-prompt behavior exactly---%
but the two inquiries we evaluate are too few
to claim construct validity at scale.
A further construct risk is that \framework measures characters removed
rather than semantic information removed:
two reductions that delete the same number of characters
can differ substantially in their effect on model behavior.
We partially mitigate this in RQ2
by attributing removals to grammar nonterminals,
which at least groups characters by their semantic role in the prompt.

\section{Conclusion}\label{sec:conclusion}
This paper presents \framework,
a blackbox minimization framework for identifying the minimal causal core of prompts.
We show that few-shot exemplars can be reduced by a mean of 65.3\%~$\pm$~15.8\%
in character count while fully preserving propositional output fidelity.
Analysis of the surviving structure reveals that models preferentially retain
the logical skeleton of exemplars
while discarding natural language prose and cross-prompt relational annotations.
Regression analysis demonstrates that model scale is the dominant factor in achievable compression,
with each additional 10B parameters corresponding to 4.3 percentage points of additional removal.
Temperature has no consistent directional effect on compression rate.
Furthermore, model architecture and training choices explain a large share
of the variance that raw parameter count cannot capture.

Our findings have three practical implications for prompt engineering:
cross-prompt annotations are redundant,
logical identifiers constitute the causal core of few-shot exemplars,
and minimization should be produced on a universal encoder and deployed on a universal decoder.
Future work will explore whether these findings generalize to other domains
and whether minimized prompts can be used for safety verification of LLM-based systems.

\bibliographystyle{IEEEtran}
\bibliography{refs}

\end{document}